# Self-consistent description of relaxation processes in systems with ultra- and deep-strong coupling

T. T. Sergeev[1,2,3,5], A. A. Zyablovsky[1,2,3,4], E. S. Andrianov[1,2,3], Yu. E. Lozovik[2,5,6]

[1]Moscow Institute of Physics and Technology, 141700, 9 Institutskiy pereulok, Moscow, Russia

[2]Dukhov Research Institute of Automatics (VNIIA), 127055, 22 Sushchevskaya, Moscow, Russia

[3]Institute for Theoretical and Applied Electromagnetics, 125412, 13 Izhorskaya, Moscow, Russia

[4]Kotelnikov Institute of Radioengineering and Electronics RAS, 125009, 11-7 Mokhovaya, Moscow, Russia

[5]Institute of Spectroscopy Russian Academy of Sciences, 108840, 5 Fizicheskaya, Troitsk, Moscow, Russia

[6]MIEM at National Research University Higher School of Economics, 123458, 34 Tallinskay, Moscow, Russia

**Abstract**

An ultra-strong coupling regime takes place in a compound system when a coupling strength between the subsystems exceeds one tenth of the system eigenfrequency. It transforms into a deep-strong coupling regime when the coupling strength exceeds the system eigenfrequency. In these regimes, there are difficulties with description of relaxation processes without explicit considering of environment degrees of freedom. To correctly evaluate the relaxation rates, it is necessary to consider the interaction of the system with its environment taking into account the counter-rotating wave and diamagnetic terms. We develop a self-consistent theory for calculation of the relaxation rates in the systems, in which the coupling strength is of the order of the system eigenfrequency. We demonstrate that the increase in the coupling strength can lead to a significant decrease in the relaxation rates. In particular, we show that for frequency-independent density of states of the environment, the relaxation rates decrease exponentially with the increase in the coupling strength. This fact can be used to suppress losses by tuning the strength coupling and the environment states.

**Introduction**

Recently, strongly coupled systems are of great interest [1-5]. The strong coupling takes place when the coupling strength between the subsystems exceeds the relaxation rates. There are a number of physical realizations of systems with strong coupling: optical [6,7], atom-cavity [7-11], polariton [12-14], optomechanical [15-17], semiconductors [18-21], superconducting [22,23] systems, etc. Further increase in the coupling strength leads to the appearance of an ultra-strong coupling (USC) and then a deep-strong coupling (DSC) regimes [24-35]. The formal definition of the ultra-strong coupling regime is that the coupling strength exceeds one tenth of the system frequency [25,27,28,31]. In this regime, the counter-rotating wave and the diamagnetic terms [31] play an important role leading to changes in the eigenenergies and the eigenstates. In particular, the state with zero number of excitations ceases to be the ground state [31].

The modification of the eigenstates leads to a change in the interaction between the system and its environment. At the same time, it is the energy exchange with an environment that causes the relaxation processes. Therefore, the transition to the ultra-strong coupling should cause the change in the relaxation processes in the system.

In this paper, we develop a model to calculate the relaxation rates in the ultra-strongly coupled system of two oscillators interacting with their own reservoirs. Within our model, we take into account the influence of the counter-rotating wave and the diamagnetic terms in the interaction Hamiltonians. We determine the dependence of the relaxation rates on the coupling strength for different densities of states in the reservoirs. We demonstrate that the change in the eigenstates occurring with the increase in the coupling strength can lead to a suppression of the relaxations in the system. In particular, we demonstrate that it is possible to achieve an exponential decrease in the relaxation rates with increasing the coupling strength. The obtained results open the new opportunities to significant reduction of losses in large variety of the strongly coupled systems.

**The model for description of relaxation**

We consider a system of two coupled oscillators, which is determined by the following Hamiltonian:

$$\hat{H}_S = \omega_0 \hat{a}_1^\dagger \hat{a}_1 + \omega_0 \hat{a}_2^\dagger \hat{a}_2 + \Omega(\hat{a}_1 + \hat{a}_1^\dagger)(\hat{a}_2 + \hat{a}_2^\dagger) + D_1(\hat{a}_1 + \hat{a}_1^\dagger)^2 + D_2(\hat{a}_2 + \hat{a}_2^\dagger)^2 \qquad (1)$$

Here $\hat{a}_{1,2}$ and $\hat{a}_{1,2}^\dagger$ are the annihilation and creation operators of the first and second oscillators, obeying the boson commutation relations $\left[\hat{a}_i, \hat{a}_j^\dagger\right] = \delta_{ij}$ [36]; $\omega_0 \hat{a}_1^\dagger \hat{a}_1$ and $\omega_0 \hat{a}_2^\dagger \hat{a}_2$ are the terms corresponding to the energies of individual oscillators, $\omega_0$ is the frequency of the oscillators ($\hbar = 1$). The term $\Omega(\hat{a}_1 + \hat{a}_1^\dagger)(\hat{a}_2 + \hat{a}_2^\dagger)$ describes the dipole-dipole interaction between the oscillators taking into account the counter-rotating wave term [31], $\Omega$ is the coupling strength between two oscillators in the system. The terms $D_1(\hat{a}_1 + \hat{a}_1^\dagger)^2$ and $D_2(\hat{a}_2 + \hat{a}_2^\dagger)^2$ are the diamagnetic terms [31]. Note that to avoid eigenenergies of the Hamiltonian tending to minus infinity, it is necessary that $D_{1,2} \geq \Omega^2 / 2\omega_0$ [31], which we use in the following consideration.

Using the Heisenberg equations [36] for the Hamiltonian (1) we obtain a closed system for the operators $\hat{a}_{1,2}$ and $\hat{a}_{1,2}^\dagger$. The derived system of equations is closed due to the boson commutation relations for all operators. Then, we pass from the equations for operators to equations for average values of the operators, and obtain the following system of equations:

$$\frac{d}{dt}\begin{pmatrix} a_1 \\ a_2 \\ a_1^* \\ a_2^* \end{pmatrix} = \begin{pmatrix} -i\omega_0 - 2iD_1 & -i\Omega & -2iD_1 & -i\Omega \\ -i\Omega & -i\omega_0 - 2iD_2 & -i\Omega & -2iD_2 \\ 2iD_1 & i\Omega & i\omega_0 + 2iD_1 & i\Omega \\ i\Omega & 2iD_2 & i\Omega & i\omega_0 + 2iD_2 \end{pmatrix} \begin{pmatrix} a_1 \\ a_2 \\ a_1^* \\ a_2^* \end{pmatrix} \qquad (2)$$

where $a_1 = \langle \hat{a}_1 \rangle$, $a_2 = \langle \hat{a}_2 \rangle$, $a_1^* = \langle \hat{a}_1^\dagger \rangle$, $a_2^* = \langle \hat{a}_2^\dagger \rangle$. The eigenstates of this system are symmetric and anti-symmetric modes, which correspond to the oscillations of the oscillators' amplitudes in and out of phase.

The interaction of the oscillators with an environment leads to an energy flow between them. When the number of degrees of freedom in the reservoirs is much greater than the one in the system, the energy predominantly flows from the system into the reservoirs. This flow results in the decrease of the energy in the system, i.e. leads to the energy relaxation.

To describe the relaxations, it is often assumed that the relaxation in each oscillator occurs independently [37,38]. In this approximation, the relaxation rates of the oscillators do not depend on the coupling strength between the oscillators and the system of equations (2) takes the following form

$$\frac{d}{dt}\begin{pmatrix} a_1 \\ a_2 \\ a_1^* \\ a_2^* \end{pmatrix} = \begin{pmatrix} -i\omega_0 - 2iD_1 - \gamma_1 & -i\Omega & -2iD_1 & -i\Omega \\ -i\Omega & -i\omega_0 - 2iD_2 - \gamma_2 & -i\Omega & -2iD_2 \\ 2iD_1 & i\Omega & i\omega_0 + 2iD_1 - \gamma_1 & i\Omega \\ i\Omega & 2iD_2 & i\Omega & i\omega_0 + 2iD_2 - \gamma_2 \end{pmatrix}\begin{pmatrix} a_1 \\ a_2 \\ a_1^* \\ a_2^* \end{pmatrix} \quad (3)$$

where $\gamma_{1,2}$ are the relaxation rates of the first and second oscillators, respectively.

Using the Equation (3) the dependence of the eigenvalues on the coupling strength can be calculated [Figure 1]. When $\gamma_1 \neq \gamma_2$, there is an exceptional point ($\Omega = \Omega_{EP}$), at which all eigenvalues coincide with each other and the eigenstates are collinear [7]. Above the exceptional point ($\Omega > \Omega_{EP}$), the real parts of all eigenvalues (i.e. the relaxation rates) are equal to each other and do not depend on the coupling strength [Figure 1a], while below the exceptional point ($\Omega < \Omega_{EP}$), the imaginary parts of the eigenvalues do not depend on the coupling strength [Figure 1b].

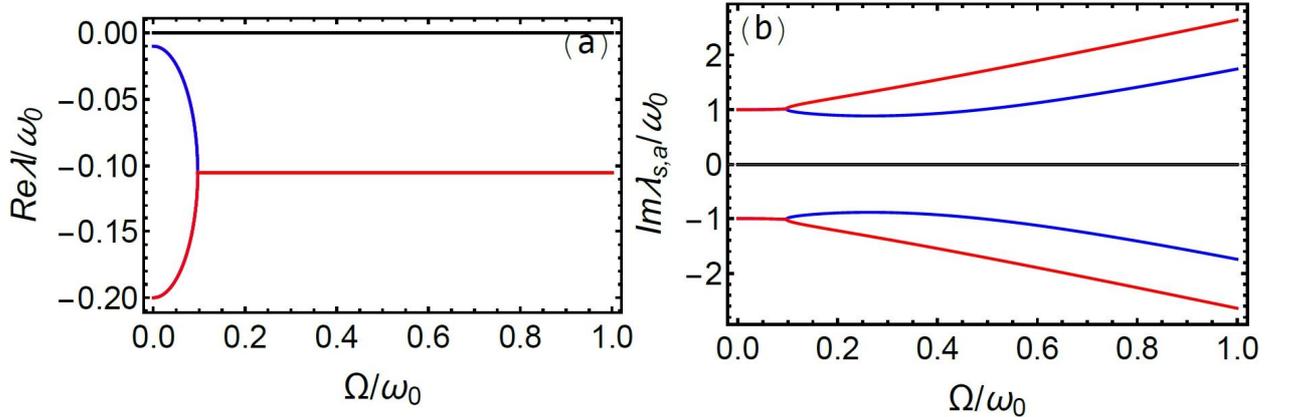

Figure 1. Dependence of the real (a) and imaginary (b) parts of the eigenvalues on the coupling strength. The eigenvalues corresponding to $a_1$ and $a_2$ are complex conjugate of the ones for $a_1^*$ and $a_2^*$. Here $\gamma_1 = 0.2\omega_0, \gamma_2 = 0.01\omega_0$.

The approach, in which the relaxation in each oscillator occurs independently, is not applicable when $\Omega$ becomes comparable with the relaxation rates [39,40]. In this case, it is necessary to take into account the change in the eigenstates of system occurring with the increase in the coupling strength [39,40]. When $\Omega$ exceeds $0.1\omega_0$, the counter-rotating wave and diamagnetic terms begin to play important role leading to modification of the eigenstates [31]. In particular, they lead to the state with zero number of excitations ceases to be the ground state [31]. This regime is called as an ultra-strong coupling regime [31].

To describe the relaxation in the ultra-strong coupling regime, we consider the total system including both the two coupled oscillators and the environment. We consider that each oscillator interacts with its own reservoir. The reservoirs are represented as sets of $N_{1,2}$ oscillators, respectively. The Hamiltonian of the total system is

$$\hat{H} = \hat{H}_S + \hat{H}_R + \hat{H}_{SR} \qquad (4)$$

Here $\hat{H}_S$ is the Hamiltonian of two coupled oscillators (see Eq. (1)). $\hat{H}_R = \sum_{k=1}^{N_1} \omega_k^{(1)} \hat{b}_k^\dagger \hat{b}_k + \sum_{k=1}^{N_2} \omega_k^{(2)} \hat{c}_k^\dagger \hat{c}_k + \sum_{k=1}^{N_1} D_b \left( \hat{b}_k + \hat{b}_k^\dagger \right)^2 + \sum_{k=1}^{N_2} D_c \left( \hat{c}_k + \hat{c}_k^\dagger \right)^2$ is the Hamiltonian of the reservoirs and $\hat{H}_{SR} = \sum_{k=1}^{N_1} g_1 (\hat{a}_1 + \hat{a}_1^\dagger)(\hat{b}_k + \hat{b}_k^\dagger) + \sum_{k=1}^{N_2} g_2 (\hat{a}_2 + \hat{a}_2^\dagger)(\hat{c}_k + \hat{c}_k^\dagger)$ is the Hamiltonian of the interaction between the oscillators and their reservoirs. $\hat{b}_k$, $\hat{c}_k$ and $\hat{b}_k^\dagger$, $\hat{c}_k^\dagger$ are the bosonic annihilation and creation operators of the oscillators in the reservoirs. $\omega_k^{(1),(2)} = \omega_0 + \delta\omega_{1,2}(k - N_{1,2}/2)$ are frequencies of the oscillators in each of the reservoirs. $g_{1,2}$ are the interaction strengths between two oscillators and their reservoirs respectively. $\sum_{k=1}^{N_1} D_b \left( \hat{b}_k + \hat{b}_k^\dagger \right)^2$ and $\sum_{k=1}^{N_2} D_c \left( \hat{c}_k + \hat{c}_k^\dagger \right)^2$ are the diamagnetic terms in the reservoirs, where $D_{b,c} = g_{1,2}^2 / 2\omega_0$.

Using the Heisenberg equations [36], we obtain a closed system for the operators $\hat{a}_{1,2}$, $\hat{a}_{1,2}^\dagger$, $\hat{b}_k$, $\hat{b}_k^\dagger$, $\hat{c}_k$, $\hat{c}_k^\dagger$. Due to the boson commutation relations for all operators the derived system of equations is closed. Then, we pass from the operators' equations to equations for average values of the operators and obtain the following system of equations:

$$\frac{da_1}{dt} = -i(\omega_0 + 2D_1)a_1 - i\Omega(a_2 + a_2^*) - 2iD_1 a_1^* - i\sum_{k=1}^{N_1} g_1(b_k + b_k^*)$$
$$\frac{da_1^*}{dt} = i(\omega_0 + 2D_1)a_1^* + i\Omega(a_2 + a_2^*) + 2iD_1 a_1 + i\sum_{k=1}^{N_1} g_1(b_k + b_k^*) \qquad (5)$$

$$\frac{da_2}{dt} = -i(\omega_0 + 2D_2)a_2 - i\Omega(a_1 + a_1^*) - 2iD_2 a_2^* - i\sum_{k=1}^{N_2} g_2(c_k + c_k^*)$$
$$\frac{da_2^*}{dt} = i(\omega_0 + 2D_2)a_2^* + i\Omega(a_1 + a_1^*) + 2iD_2 a_2 + i\sum_{k=1}^{N_2} g_2(c_k + c_k^*) \qquad (6)$$

$$\frac{db_k}{dt} = -i(\omega_k^{(1)} + 2D_b)b_k - 2iD_b b_k^* - i g_1(a_1 + a_1^*)$$
$$\frac{db_k^*}{dt} = i(\omega_k^{(1)} + 2D_b)b_k^* + 2iD_b b_k + i g_1(a_1 + a_1^*) \qquad (7)$$

$$\frac{dc_k}{dt} = -i(\omega_k^{(2)} + 2D_c)c_k - 2iD_c c_k^* - i g_2(a_2 + a_2^*)$$
$$\frac{dc_k^*}{dt} = i(\omega_k^{(2)} + 2D_c)c_k^* + 2iD_c c_k + i g_2(a_2 + a_2^*) \qquad (8)$$

where $a_1 = \langle \hat{a}_1 \rangle$, $a_2 = \langle \hat{a}_2 \rangle$, $a_1^* = \langle \hat{a}_1^\dagger \rangle$, $a_2^* = \langle \hat{a}_2^\dagger \rangle$, $b_k = \langle \hat{b}_k \rangle$, $b_k^* = \langle \hat{b}_k^\dagger \rangle$, $c_k = \langle \hat{c}_k \rangle$ and $c_k^* = \langle \hat{c}_k^\dagger \rangle$.

### The relaxations in ultra-strong and deep-strong coupling regimes

We use the Equations (5)-(8) to calculate the relaxation rates in the system for the different coupling strengths. In the case of the reservoirs with a finite number of degrees of freedom, the collapses and revivals of energy are observed in the oscillators' dynamics [Figure 2] [41]. The time of the first appearance of revival is $T_R = 2\pi / \delta\omega$ [41]. At $t < T_R$ the temporal dynamics of the oscillators' amplitudes are approximated by a sum of two exponents. In the limit of infinite number of the degrees of freedom in the reservoirs, one has $T_R \to \infty$ and the evolution of the first and second oscillators obtained from the Eqns. (5)-(8) corresponds to the one in the open system described by Eqns. (3). In this limit, the factors of exponents correspond to the real parts of eigenvalues of the non-Hermitian system of two coupled oscillators.

To calculate the relaxation rates in the system of coupled oscillators, we simulate the Hermitian system with a large number of degrees of freedom in the reservoirs. Then, we use the Fourier transform over time interval $t \in [0, T_R]$ to find the relaxation rates. Our calculations show that in the system spectrum, there are two peaks at the positive frequencies (and also two peaks at the negative frequencies), which correspond to oscillations in and out of phase. These oscillations can be associated with the symmetric and anti-symmetric modes of the Eq. (3).

We find that when $\Omega \ll \omega_0$, the system behavior at $t < T_R$ calculated by the Eqns. (5)-(8) is similar to the one calculated by the Equation with the relaxation (3) [Figure 2a]. In this limit, the behavior of the relaxation rates of the Eq. (3) coincides with the one predicted from the Eqns. (5)-(8).

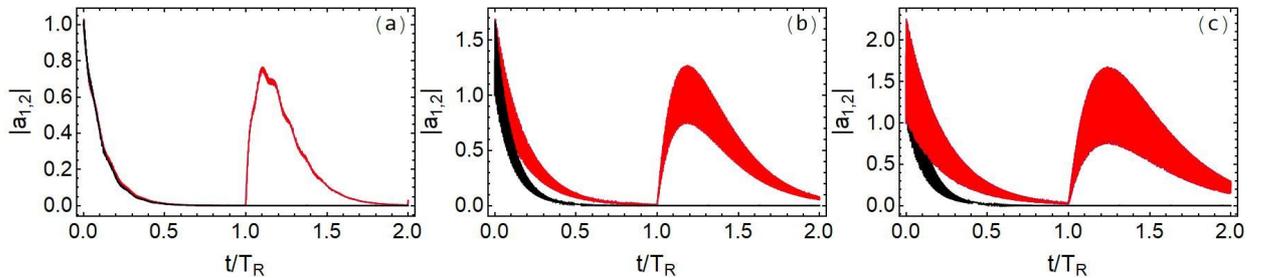

Figure 2. Dependence of the amplitudes of first oscillators calculated by Eqns. (5)-(8) (the red lines) and by Eq. (3) (the black lines) systems on time with three different coupling strengths: (a) $\Omega = 0.05\omega_0$; (b) $\Omega = 0.5\omega_0$; (c) $\Omega = 0.8\omega_0$. In the Eqns. (5)-(8) $\delta\omega = 0.01\omega_0$, $T_R = 200\pi\omega_0^{-1}$ and number of oscillators in the reservoirs $N = 300$. In the Eq. (3) $\gamma_1 = 0.02\omega_0$, $\gamma_2 = 0.01\omega_0$. The behavior of the second oscillator is the same.

At the increase in the coupling strength, significant difference in the behavior of the relaxation rates from the predictions based on the Eq. (3) appears. This difference is that the relaxation rates cease to be independent on the coupling strength [cf. Figures 1a and 3]. The relaxation rates of the symmetric and anti-symmetric modes have different dependence on the coupling strength [Figures 3]. However, when $\Omega$ tends to the oscillators' frequency $\omega_0$, the dependence of the relaxation rates of both modes on $\Omega$ is well approximated by a function $\gamma(\Omega) = \gamma(\Omega = 0)\exp(-\Omega/\omega_0)$ [Figure 3]. Thus, it is seen that the increase in the coupling strength leads to a decrease of the relaxation rates in the system.

Note that our model takes into account the counter-rotating wave and diamagnetic terms in the Hamiltonian and therefore it is suitable for the description of the system dynamics in the ultra-strong coupling ($\Omega > 0.1\omega_0$) and the deep-strong coupling ($\Omega > \omega_0$) regimes. Therefore, we can conclude that the transition to ultra-strong and deep-strong coupling regimes enable to decreases the influence of the environment on the system.

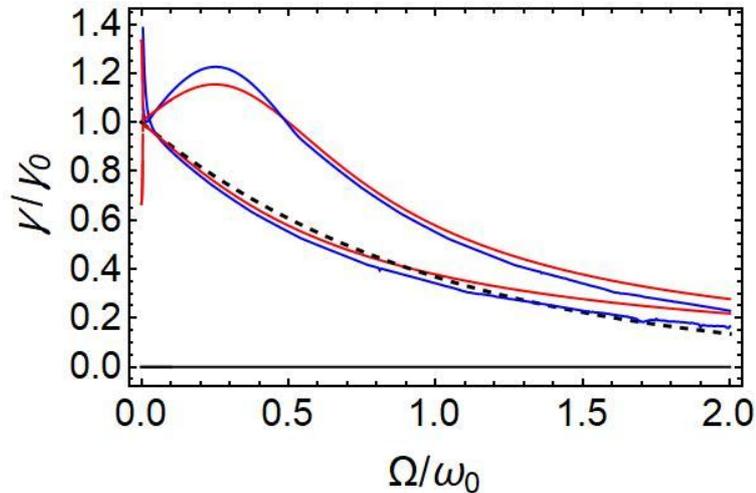

Figure 3. Dependence of the relaxation rates of the system states on coupling strength. The blue lines are calculated by the Eqns. (5)-(8). The red lines are calculated by the analytical expression (16). The dashed black line is an approximation of the obtained curves by the function $\gamma(\Omega) = \gamma(\Omega=0)\exp(-\Omega/\omega_0)$.

It is worth noting here that the system is non-symmetric, that is the interaction strength of the first oscillators with its reservoir differs from the one of the second oscillators with its reservoir ($g_1 \neq g_2$). Therefore, there is a possibility that the decrease in the relaxation rates is associated with some rearrangement of the eigenstates. To prove that the decrease in relaxation rates is not related to the non-symmetric interaction with the reservoirs, we consider the symmetrical system, in which the interactions of the oscillators with their reservoirs are the same ($g_1 = g_2$). However, even in this case, the decrease in the relaxation rates takes place with the increase in the coupling strength [Figure 4].

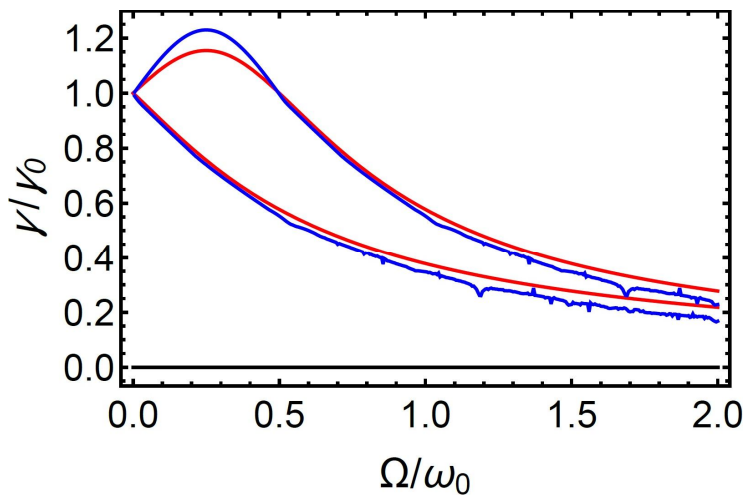

Figure 4. Dependence of the relaxation rates of the system states on coupling strength in the system with the same reservoirs. The blue lines are calculated by the Equations (5)-(8). The red lines are calculated by the analytical expression (16).

**The analytical model**

To derive analytical expressions for the relaxation rates, we use the approach developed in theory of environment-assisted strong coupling (EASC) regime [39]. Within this approach, we find the eigenstates and the eigenfrequencies of two coupled oscillators and then eliminate from consideration the reservoirs' degrees of freedom. The interaction of the system with the reservoirs occurs at the eigenfrequencies [39], which differ from the frequencies of the individual oscillators and depend on $\Omega$. This fact leads to the dependence of the relaxation rates on the coupling strength.

At the beginning, we rewrite the Hamiltonian (4) for the oscillator's system through the operators of coordinate and momentum. For first and second oscillators, the operators of the coordinates and the momentums are given as $\hat{x}_{1,2} = \frac{1}{\sqrt{2\omega_0}}(\hat{a}_{1,2} + \hat{a}^\dagger_{1,2})$ and $\hat{p}_{1,2} = -i\sqrt{\frac{\omega_0}{2}}\left(\hat{a}_{1,2} - \hat{a}^\dagger_{1,2}\right)$ [36], while for the oscillators in the reservoirs, the operators of the coordinates and the momentums are $\hat{y}^{(1)}_k = \frac{1}{\sqrt{2\omega_k}}(\hat{b}_k + \hat{b}^\dagger_k)$, $\hat{y}^{(2)}_k = \frac{1}{\sqrt{2\omega_k}}(\hat{c}_k + \hat{c}^\dagger_k)$ and $\hat{q}^{(1)}_k = -i\sqrt{\frac{\omega_k}{2}}\left(\hat{b}_k - \hat{b}^\dagger_k\right)$, $\hat{q}^{(2)}_k = -i\sqrt{\frac{\omega_k}{2}}\left(\hat{c}_k - \hat{c}^\dagger_k\right)$, respectively. These operators obey the standard commutation relations for the coordinates and the momentums [36]

In new variables, the Hamiltonian has the form:

$$\hat{H} = \hat{H}_S + \hat{H}_R + \hat{H}_{SR} \qquad (9)$$

where the Hamiltonian of two coupled oscillators is

$$\hat{H}_S = \frac{\hat{p}_1^2}{2} + \frac{\hat{p}_2^2}{2} + \frac{\omega_o^2 \hat{x}_1^2}{2} + \frac{\omega_o^2 \hat{x}_2^2}{2} + 2\Omega\omega_0 \hat{x}_1 \hat{x}_2 + 2D_1\omega_0 \hat{x}_1^2 + 2D_2\omega_0 \hat{x}_2^2, \qquad (10)$$

the Hamiltonian of the reservoirs is

$$\hat{H}_R = \sum_k \left( \frac{\hat{q}^{(1)2}_k}{2} + \frac{\hat{q}^{(2)2}_k}{2} + \frac{\omega^{(1)2}_k \hat{y}^{(1)2}_k}{2} + \frac{\omega^{(2)2}_k \hat{y}^{(2)2}_k}{2} + 2D_b \omega^{(1)}_k \hat{y}^{(1)2}_k + 2D_c \omega^{(2)}_k \hat{y}^{(2)2}_k \right), \qquad (11)$$

and the Hamiltonian of the interaction between the system of two coupled oscillators and the reservoirs is

$$\hat{H}_{SR} = 2\sum_k \left( \sqrt{\omega_0 \omega^{(1)}_k} g_1 \hat{x}_1 \hat{y}^{(1)}_k + \sqrt{\omega_0 \omega^{(2)}_k} g_2 \hat{x}_2 \hat{y}^{(2)}_k \right). \qquad (12)$$

Note that such a representation contains the counter-rotating wave and diamagnetic terms, which makes it possible to describe analytically the relaxation processes in the systems with ultra-strong and deep-strong coupling.

Using the Heisenberg equations [36], for the momentum operators ($\frac{d\hat{p}_1}{dt} = i\left[\hat{H}, \hat{p}_1\right]$, ...), we obtain the following system of equations

$$\hat{\ddot{p}}_1 + \left(\omega_0^2 + 4D_1\omega_0\right)\hat{x}_1 = -2\Omega\omega_0 \hat{x}_2 - 2\sum_k \sqrt{\omega_0 \omega^{(1)}_k} g_1 \hat{y}^{(1)}_k$$

$$\hat{\dot{p}}_2 + \left(\omega_0^2 + 4D_2\omega_0\right)\hat{x}_2 = -2\Omega\omega_0\hat{x}_1 - 2\sum_k \sqrt{\omega_0\omega_k^{(2)}}\, g_2 \hat{y}_k^{(2)} \qquad (13)$$

$$\hat{\dot{q}}_k^{(1)} + \left(\omega_k^{(1)2} + 4D_b\omega_k^{(1)}\right)\hat{y}_k^{(1)} = -2\sqrt{\omega_0\omega_k^{(1)}}\, g_1 \hat{x}_1$$

$$\hat{\dot{q}}_k^{(2)} + \left(\omega_k^{(2)2} + 4D_c\omega_k^{(2)}\right)\hat{y}_k^{(2)} = -2\sqrt{\omega_0\omega_k^{(2)}}\, g_2 \hat{x}_2$$

Passing from the operators' equations to the equations for the averages ($x_{1,2} = \langle \hat{x}_{1,2} \rangle$, $y_k^{(1),(2)} = \langle \hat{y}_k^{(1),(2)} \rangle$, $\ddot{x}_{1,2} = \langle \hat{\dot{p}}_{1,2} \rangle$, $\ddot{y}_k^{(1),(2)} = \langle \hat{\dot{q}}_k^{(1),(2)} \rangle$), we obtain the equations describing the dynamics of a classical system:

$$\ddot{x}_1 + \left(\omega_0^2 + 4D_1\omega_0\right)x_1 = -2\Omega\omega_0 x_2 - \sum_k \tilde{g}_1 y_k^{(1)}$$

$$\ddot{x}_2 + \left(\omega_0^2 + 4D_2\omega_0\right)x_2 = -2\Omega\omega_0 x_1 - \sum_k \tilde{g}_2 y_k^{(2)}$$

$$\ddot{y}_k^{(1)} + \left(\omega_k^{(1)2} + 4D_b\omega_k^{(1)}\right)y_k^{(1)} = -\tilde{g}_1 x_1 \qquad (14)$$

$$\ddot{y}_k^{(2)} + \left(\omega_k^{(2)2} + 4D_c\omega_k^{(2)}\right)y_k^{(2)} = -\tilde{g}_2 x_2$$

where $\tilde{g}_1 = 2\sqrt{\omega_0\omega_k^{(1)}}\, g_1$ and $\tilde{g}_2 = 2\sqrt{\omega_0\omega_k^{(2)}}\, g_2$ are the effective coupling strengths of the first and second oscillators with its reservoirs.

To move further, we will assume that $D_1 = D_2 = \Omega^2/\omega_0$, $\omega_k^{(1)} = \omega_k^{(2)} = \omega_k$ and $D_b = D_c = 0$. In this case, the equations (14) break down into equations for the symmetric and anti-symmetric modes $x_{s,a} = \dfrac{x_1 \pm x_2}{\sqrt{2}}$, $y_k^{(s,a)} = \dfrac{\tilde{g}_2 y_k^{(1)} \pm \tilde{g}_1 y_k^{(2)}}{\sqrt{\tilde{g}_1^2 + \tilde{g}_2^2}}$ with frequencies $\omega_{s,a} = \sqrt{\omega_0^2 \pm 2\Omega\omega_0 + 4\Omega^2}$.

Following [39], we eliminate the reservoirs' degrees of freedom within the Born-Markovian approximation [37,38] and then move back to the equations for the amplitudes of the individual oscillators $a_{1,2} = \dfrac{\omega_0 x_{1,2} + ip_{1,2}}{\sqrt{2\omega_0}}$. As a result, we obtain the following equations:

$$\dot{a}_1 = -\left(\frac{\beta_1^{(1)} + \beta_2^{(2)}}{2} + \frac{\beta_2^{(1)} + \beta_1^{(2)}}{2}\right)a_1 - \left(\frac{\beta_1^{(1)} - \beta_2^{(2)}}{2} - \frac{\beta_2^{(1)} - \beta_1^{(2)}}{2} + i\frac{\omega_s - \omega_a}{2}\right)a_2$$

$$\dot{a}_2 = -\left(\frac{\beta_1^{(1)} - \beta_2^{(2)}}{2} + \frac{\beta_2^{(1)} - \beta_1^{(2)}}{2} + i\frac{\omega_s - \omega_a}{2}\right)a_1 - \left(\frac{\beta_1^{(1)} + \beta_2^{(2)}}{2} - \frac{\beta_2^{(1)} + \beta_1^{(2)}}{2}\right)a_2 \qquad (15)$$

Here $\beta_1^{(1)} = \Gamma^{(+)}(\omega_s)/8\omega_s^2$, $\beta_2^{(2)} = \Gamma^{(+)}(\omega_a)/8\omega_a^2$, $\beta_2^{(1)} = \Gamma^{(-)}(\omega_a)/(8\omega_s\omega_a)$, $\beta_1^{(2)} = \Gamma^{(-)}(\omega_s)/(8\omega_s\omega_a)$, where $\Gamma^{(\pm)}(\omega_{s,a}) = \sum_{\omega_k} \pi\left(\rho^{(1)}\tilde{g}_1^2 \pm \rho^{(2)}\tilde{g}_2^2\right)\delta(\omega_k - \omega_{s,a})$ and $\rho^{(1),(2)}(\omega)$ are the densities of the reservoirs' states. Note that we have excluded from consideration the terms responsible for the frequency shift [39].

In the case of frequency-independent densities of states in the reservoirs, after integrating over frequency [37,38], we obtain that $\Gamma^{(\pm)}(\omega_{s,a}) = 4\omega_0\omega_{s,a}\left(\gamma_1(\omega_{s,a}) \pm \gamma_2(\omega_{s,a})\right)$, where

$\gamma_{1,2}(\omega_{s,a}) = \dfrac{\pi g_{1,2}^2(\omega_{s,a})}{\delta \omega_{1,2}} \rho^{(1),(2)}(\omega_{s,a})$ are the relaxation rates derived in the rotating-wave approximation [37,38] and $\omega_{s,a} = \sqrt{\omega_0^2 \pm 2\Omega\omega_0 + 4\Omega^2}$.

The real parts of the eigenvalues of the Eq. (15) determine the relaxation rates of the eigenstates of the system with the Hamiltonian (9). Calculating the eigenvalues of (15), we obtain the following expression for the relaxation rates:

$$\lambda_{s,a} = -\dfrac{\omega_s^2 \Gamma_a^{(+)} + \omega_a^2 \Gamma_s^{(+)} \mp \sqrt{4\omega_s^2 \omega_a^2 \Gamma_s^{(-)} \Gamma_a^{(-)} + \left(\omega_s^2 \left(8\omega_a^2(\omega_s - \omega_a) + i\Gamma_a^{(+)}\right) - i\omega_a^2 \Gamma_s^{(+)}\right)^2}}{16 \omega_s^2 \omega_a^2} \qquad (16)$$

where $\Gamma_{s,a}^{(\pm)} = 4\omega_0 \omega_{s,a} \left(\gamma_1(\omega_{s,a}) \pm \gamma_2(\omega_{s,a})\right)$.

The analytical expressions (16) are in good agreement with the results of numerical calculations based on the Eqns. (5)-(8) [cf. blue and red lines in Figures 3 and 4]. Note that the expression (16) are derived in the assumption that the diamagnetic terms in the reservoirs, $D_b = D_c = 0$. In the total Equations (5)-(8), the diamagnetic terms $D_{b,c} = g_{1,2}^2 / 2\omega_0$ and their influence can be significant even in the case of non-coupled system ($\Omega = 0$). This fact can be a reason of the existed discrepancy between the predictions of the analytical model and the numerical simulation [cf. the red and blue lines in Figures 3 and 4].

**The dependence of the relaxation rates on the coupling strength at different frequency dispersion of $g_{1,2}(\omega)$**

In the previous sections, we consider the case when the densities of states in the reservoirs and the coupling strengths of the oscillators with its reservoirs ($g_{1,2}$) are frequency-independent. In this case, the relaxation rates $\gamma$ decrease with the increase in the coupling strength [Figures 3 and 4] and $\gamma \sim \omega^{-1}$ (see expression (16) taking into account that $\gamma = -\operatorname{Re}\lambda$).

In this section, we study the behavior of relaxation rates, $\gamma$, on $\Omega$ for different frequency dependencies of the coupling strengths of the oscillators with its reservoirs, $g_{1,2}(\omega)$. To be specific, we consider two cases: $g_{1,2} \sim \omega$ and $g_{1,2} \sim \omega^{5/4}$. Using the Eq. (16), it is easy to derive that in these cases, the relaxation rates depend on the frequency as $\gamma \sim \omega$ and $\gamma \sim \omega^{3/2}$, respectively (see expression (16)).

Using the Eqns. (5)-(8) and the method described in the section "The relaxations in ultra-strong and deep-strong coupling regimes", we calculate the dependence of the relaxation rates on $\Omega$. Our calculations show that for both frequency dispersions ($g_{1,2} \sim \omega$ and $g_{1,2} \sim \omega^{5/4}$) the increase in the coupling strength, $\Omega$, leads to the increase in the relaxation rates [Figure 5]. This behavior qualitatively differs from the one in the case of the frequency-independent dispersion ($g_{1,2}(\omega) = const$), which has been considered early [cf. Figures 3 and 5].

Thus, the analytical model developed by us confirms that depending on the frequency dispersions, $g_{1,2}(\omega)$, the increase in the coupling strength, $\Omega$, can lead to both the decrease [see red lines in Figures 3 and 4] and the increase [see red lines in Figure 5] in the relaxation rates.

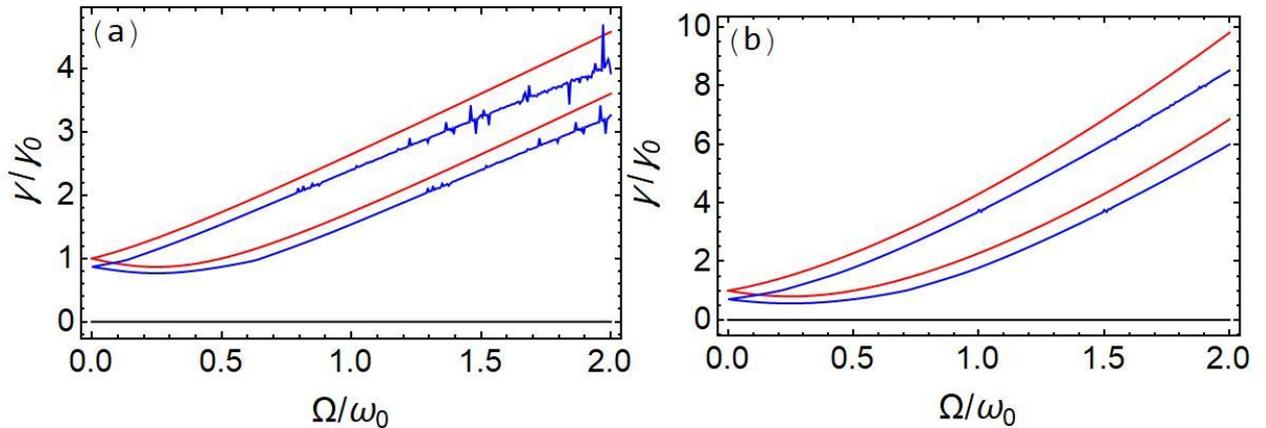

Figure 5. Dependence of the relaxation rates of the system states on coupling strength when $g_{1,2} \sim \omega$ (a) and $g_{1,2} \sim \omega^{5/4}$ (b). The blue lines are calculated by the Eqns. (5)-(8). The red lines are calculated by the analytical expression (16).

**Conclusions**

We develop an approach to determine the relaxation rates in the systems with ultra-strong and deep-strong coupling regimes. This approach enables to take into account an influence of the counter-rotating wave and the diamagnetic terms in the Hamiltonian on the relaxation processes. We determine the dependence of the relaxation rates on the coupling strength for the different types of the reservoirs. We demonstrate that the increase in the coupling strength can be accompanied by both the decrease and the increase in the relaxation rates. We demonstrate that for the frequency-independent densities of states in the reservoirs, the exponential decrease in the relaxation rates with the coupling strength takes place. Our results demonstrate significant influence of the counter-rotating wave and the diamagnetic terms on the relaxation processes in the system with ultra-strong and deep-strong coupling regimes.

**Acknowledgements**

The study was financially supported by a Grant from Russian Science Foundation (project No. 23-42-10010). T.T.S. and Yu.E.L. thank foundation for the advancement of theoretical physics and mathematics "Basis". Yu.E.L. acknowledged Basic Research Program at the National Research University HSE.